\documentclass[journal,10pt,twocolumn]{IEEEtran}
\usepackage{epsfig}
\usepackage{graphicx}
\usepackage{subfigure}
\usepackage{algorithm}
\usepackage{algorithmic}
\usepackage{color}

\newtheorem{theorem}{Theorem}

\newtheorem{definition}{Definition}

\newcommand{\kFPTAS}	{{\sf FPTAS}}

\usepackage{amssymb}
\usepackage{makeidx}
\usepackage{amsmath}
\usepackage{graphicx}
\usepackage{epsf}
\usepackage{epsfig}
\usepackage{psfig}
\usepackage{ccaption}
\usepackage{array}
\usepackage{tabularx}
\usepackage{multirow}
\usepackage{epsfig}
\usepackage{cite}
\begin{document}
%

\title{QoS Routing in Smart Grid}
%
\author{Husheng Li and Weiyi Zhang
\thanks{H. Li is with the Department of Electrical Engineering
and Computer Science, the University of Tennessee, Knoxville, TN,
37996 (email: husheng@eecs.utk.edu). W. Zhang is with the Department of Computer Science, North Dakota State University, Fargo, ND, 58105. This work was supported by the National Science
Foundation under grants CCF-0830451 and
ECCS-0901425.}}

\maketitle
\begin{abstract}
Smart grid is an emerging technology which is able to control the power load via price signaling. The communication between the power supplier and power customers is a key issue in smart grid.
Performance degradation like delay or outage may cause significant impact on the stability of the pricing based control and thus the reward of smart grid. Therefore, a QoS mechanism is proposed for the communication system in smart grid, which incorporates the derivation of QoS requirement and applies QoS routing in the communication network. For deriving the QoS requirement, the dynamics of power load and the load-price mapping are studied. The corresponding impacts of different QoS metrics like delay are analyzed. Then, the QoS is derived via an optimization problem that maximizes the total revenue. Based on the derived QoS requirement, a simple greedy QoS routing algorithm is proposed for the requirement of high speed routing in smart grid. It is also proven that the proposed greedy algorithm is a $K$-approximation. Numerical simulation shows that the proposed mechanism and algorithm can effectively derive and secure the communication QoS in smart grid.
\end{abstract}
\section{Introduction}
In recent years, power grids are experiencing a revolutionary technological transformation. One significant feature is that electric appliances can receive realtime power price via communication networks and optimize its power consumption level according to the current power price. Then, the power utilization efficiency is significantly improved and the global energy consumption is reduced to combat the crisis of energy resource.

In smart grid, a key challenge is how to adapt the communication network to the context of power price transmission. Obviously, the data flow of power price cannot be elastic since it should be realtime; otherwise, it may incur a significant loss if the expired power price is used. Therefore, the data transmission of power price must be equipped with quality of service (QoS) guarantee. This incurs two important questions unique to smart grid:
\begin{itemize}
\item {\em How to define the QoS requirement in the context of smart grid?}
\item {\em How to ensure the QoS requirement from the home appliance in the communication network?}
\end{itemize}
\begin{figure}[ht!]
  \centering
  \includegraphics[scale=0.4]{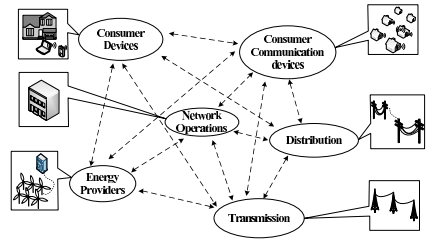}
  \caption{The network perspective of smart grid}
  \label{fig:grid}
  \vspace{-0.1in}
\end{figure}
In this paper, we {\em answer the above two questions by proposing a QoS system for smart grid.} The proposed QoS framework plays the role of interface between the power market and the communication networks. Once a set of reasonable QoS metrics can be derived in the context of smart grid, many QoS ensuing approaches can be applied to guarantee the performance gain introduced by the technology of smart grid.

To answer the first question, we need to study the detailed mechanism of power price. Take video streaming for instance. To propose a QoS requirement for video streaming, the source coder must be aware of the impacts of different factors like delay or jitter on the video quality and then derive a suitable QoS requirement. Therefore, we study the impact of QoS parameters on the reward of home appliance. For simplicity, we study only two QoS parameters, namely the delay and outage probability. The framework proposed in this paper also applies to many other QoS metrics. We first introduce the mechanism of power price based on the dynamics of load. Then, we build a reward system for the home appliance based on the power price and the utility function of the appliance, thus obtaining the impact of delay and outage on the reward of home appliance. Finally, the QoS requirement is derived by optimizing the reward.

To answer the second question, we focus on routing methodologies meeting the derived QoS requirement. We focus on providing multiple QoS-aware routing within multiple (more than 2) constraints.
Given the heterogeneity of the smart grid, traditional schemes, such as fully polynomial-time approximations \cite{XueSenZha07}\cite{Lorenz01}\cite{XueZhaTanThu08}, cannot be directly applied due to the requirements of high computing and storage capabilities.
An efficient, which can be implemented by both powerful and resource-limited devices, and effective, which provides provably-good performance, algorithm is needed for the QoS routing in smart grid.
In this part, we present a simple greedy algorithm for the multi-constrained QoS routing.
Moreover, we prove that our greedy algorithm is a $K$-approximation ($K$ is the number of constraints).
In addition, our solution can be implemented in a distributed manner.

The remainder of this paper is organized as follows.
The system model is introduced in Section \ref{sec:model}. Then, the QoS requirement is derived in Section \ref{sec:derivation} while the QoS routing is discussed in Section \ref{sec:routing}. Numerical results and conclusions are provided in Sections \ref{sec:numerical} and \ref{sec:conclusion}, respectively.

\section{System Model}\label{sec:model}
We consider a simplified model for smart grid by considering only the QoS requirement in the power price inquiry. We assume that a home appliance receives power price from the power market. A QoS requirement is sent from the home appliance to the control center of the communication network. Then, the control center assigns one or more route for the home appliance to guarantee the QoS requirement. Smart devices, such as smart meter, and electricity generator can be viewed as the nodes throughout a network. All the transmission medium, such as fiber, wireless, broadband over power line, WiMax, GPRS, Ethernet, and radio, form the links in a network.
As shown in Fig. \ref{fig:grid}, the whole infrastructure of smart grid can be represented by a communication network structure, which is designed to optimize a smart grid investment.

It is worth noting that smart gird is a heterogeneous network. Various electric equipments, with dramatically different resource limits, such as computing power, storage capability, are integrated in the grid.
Meanwhile, wireless network technology is utilized in combination with a utilities fiber or Ethernet communications infrastructure.
To provide QoS-aware routing for smart grid, we must consider the heterogeneity of the network and provide solutions that could be applied for all the devices in the networks.

\section{Derivation of QoS Requirement}\label{sec:derivation}
A QoS requirement usually includes specifications like average delay, jitter and connection outage probability. To derive the QoS requirement, the following problems should be addressed in the study:
\begin{itemize}
\item How to describe the probabilistic dynamics of the power system?
\item How to evaluate the impact of different QoS specifications on the smart grid system? For example, how does a long communication delay affect the system performance?
\item How to derive QoS requirement due to the corresponding impact?
\end{itemize}
 In this section, we provide an approach to address the above three key questions and thus derive the QoS requirements for delay and outage probability.

\subsection{Probabilistic Dynamics of the Power System}

\begin{figure}
  \centering
  \includegraphics[scale=0.4]{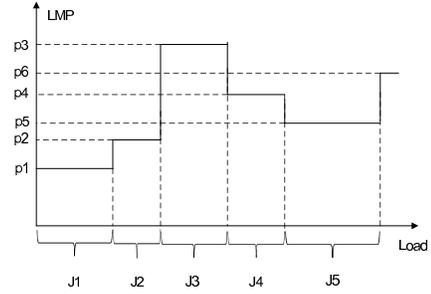}
  \caption{An illustration of the mapping between load and LMP.}\label{fig:mapping}
\end{figure}

Power price is typically determined by locational margin price (LMP) \cite{Bo2009} driven by the load which varies with time. A constrained optimization problem can be used to derive the LMP from the load and other parameters, where the Lagrange factors of the constraints are considered as prices \cite{Stoft2002}. In practical systems, we can use a piecewise curve, as illustrated in Fig. \ref{fig:mapping}, to accomplish the mapping between the load and the price. Note that, we have finite numbers of prices, denoted by $Q$, in Fig. \ref{fig:mapping}. Therefore, we denote by $q_1$, $q_2$, ..., $q_Q$ these prices. The intervals of loads corresponding to the prices $q_1$, ..., $q_Q$ are denoted by $J_1$, ..., $J_Q$, respectively. We assumed that the load is uniformly distributed within the corresponding interval given the price.

The load is random due to many random factors like the power generation and consumption level. We can model it as the positive part of a Gaussian random variable, whose probability density function (PDF) is given by
\begin{eqnarray}\label{eq:evolution}
f(D_t)=\frac{\exp\left(-\frac{(D_t-\mu_t)^2}{2\sigma_t}\right)}{\int_0^{D_{\max}} \exp\left(-\frac{(y-\mu_t)^2}{2\sigma_t}\right)dy},
\end{eqnarray}
where $D_t$ is the load at time slot $t$, $D_{\max}$ is the maximal possible load, $\mu_t$ and $\sigma_t$ are the expectation and variance.

Then, we model the Gaussian distribution parameters as functions of the elapsed time. Suppose that, at time slot 0, the true value of the load is $D_0$. Then, we assume that the load distribution at time slot $t$ satisfies that following laws:
\begin{itemize}
\item The expectation $\mu_t$ of the Gaussian distribution is equal to $D_0$. The rationale is that the prediction should be unbiased.

\item The variance $\sigma_t$ satisfies $\sigma_t=\theta t$, where $\theta$ is a parameter and can be estimated from measurements, i.e. the variance increases linearly with the elapsed time, which is similar to a Brownian motion.
\end{itemize}

\subsection{Impact of Delay}

At time slot $t$, the power price and power consumption are denoted by $p_t$ and $x_t$, respectively. We assume a time-invariant utility function for the power consumption and denote it by $U(x_t)$. The decision of power consumption is based on the known power price, which means that $x_t$ is a function of $p_{\tau_t}$. For simplicity, we assume that the optimal power consumption level maximizes the following metric:
\begin{eqnarray}
x_t(p)=\arg\max_{x}\left(U(x)-px\right),
\end{eqnarray}
where $p$ is the price adopted by the home appliance. It could be different from the true value due to delay. We assume that $U$ is an increasing, strictly concave and continuously differentiable function. We also assume that the first order derivative of $U$, denoted by $\dot{U}$, ranges from $\infty$ to $0$. Based on these assumptions, the optimal power consumption level is thus given by
$x_t(p)=\dot{U}^{-1}(p),$
which is derived from the first order condition of optimality, i.e. $\dot{U}(x)-p=0$.

Suppose that the communication delay is $d$ time slots. Then, at time slot $t$, the price used for optimizing the power consumption level is $p_{t-d}$. Hence, the cost incurred by the communication delay, as a function of the delay, is given by
\begin{eqnarray}
C(d)&=&E\left[U(x(p_t))-p_tx(p_t)\right.\nonumber\\
&-&\left.\left(U(x(p_{t-d}))-p_tx(p_{t-d})\right)\right],
\end{eqnarray}
where the expectation is over all realizations of the power price and can be computed using the probabilistic dynamics of the power price discussed in Section \ref{sec:derivation}. A.

\subsection{Impact of Outage}
It is also possible that the communication link experiences an outage such that the home appliance cannot obtain the realtime power price. In such a situation, the home appliance can only use a default power price, which is independent of the time. We assume that the default power price equals the average power price, which is denoted by $\bar{p}$. Then, the expected loss incurred by the outage is given by
\begin{eqnarray}
L(\zeta)=\zeta E\left[U(x(p_t))-p_tx(p_t)-\left(U(x(\bar{p}))-p_tx(\bar{p})\right)\right],
\end{eqnarray}
where $\zeta$ is the outage probability.

\subsection{Derivation of QoS Requirement}
If there is no constraint on the delay, the delay requirement of the home appliance should be as low as possible. However, it is expensive for the network to achieve a very low communication delay. Therefore, the system can control the delay requirement using a delay dependent price, namely $P(d)$. Then, the delay requirement of the home appliance is to minimize the total cost, i.e. the average loss incurred by using the old price and the price taxed by the communication network. The optimal delay requirement is then given by
\begin{eqnarray}
d^*=\arg\min_d \left(C(d)+P(d)\right).
\end{eqnarray}

Similar approach can be applied for deriving the requirement of outage probability. Suppose that there is a tax for the communication with outage probability $\zeta$, which is denoted by $T(\zeta)$. Then, the optimal requirement of outage probability is given by
\begin{eqnarray}
\zeta^*=\arg\min_\zeta \left(\zeta L(\zeta)+T(\zeta)\right).
\end{eqnarray}

When the QoS specification includes both delay and outage probability, the optimal QoS requirement is then given by
\begin{eqnarray}\label{eq:joint}
(d^*,\zeta^*)=\arg\min_{\lambda, \zeta} (1-\zeta)C(d)+\zeta L(\zeta)+P(d)+T(\zeta).
\end{eqnarray}

\section{QoS Routing Algorithm}\label{sec:routing}

After deriving the QoS requirements, we will study how to deliver transmission in smart grid with multi-constrained QoS routing problems with $K \geq 2$ additive QoS parameters.
%
%
%
%
%
%
\subsection{MCR Problem}
A smart grid is modeled by an edge weighted directed graph $G = (V, E, \omega$), where $V$ is the set of \emph{n nodes}, including end users, smart meters and other electric devices, \emph{E} is the set of \emph{m edges}, and $\omega = (\omega_1, ... , \omega_K$) is an \emph{edge weight vector} so that $\omega_k(e) \geq$ \emph{0} is the $k^{th}$ \emph{weight} of edge $e$.
For a path $p$ in $G$, the $k^{th}$ \emph{weight} of $p$, denoted by $\omega_k(p)$, is the sum of the $k^{th}$ weights over the edges on $p$: $\omega_k(p) = \sum\limits_{e \in p} \omega_k(e)$.

\begin{definition} [Multi-Constrained Routing ({\sf MCR})]
\label{def:MCR}
Given an edge weighted directed graph $G = (V, E, \omega)$, with $K$ nonnegative real-valued edge weights $\omega_k(e)$ associated with each edge $e$, a constraint vector $W = (W_1, ... , W_K)$ where each $W_k$ is a positive constant; and a source-destination node pair $(s, t)$. The {\sf MCR} problem seeks an $s$ {$\rightarrow$} $t$ path $p$ such that $\omega_k(p) \leq W_k, 1 \leq k \leq K$.
\hfill $\Box$
\end{definition}

The inequality $\omega_k(p) \leq W_k$ is called \emph{the $k^{th}$ QoS constraint}.
A path $p$ satisfying all $K$ QoS constraints is called a \emph{feasible path or a feasible solution} of {\sf MCR} problem.
An {\sf MCR} problem is said to be \emph{feasible} if it has a feasible path, and \emph{infeasible} otherwise.
%

\begin{figure}
\center
\includegraphics[width=2.50in]{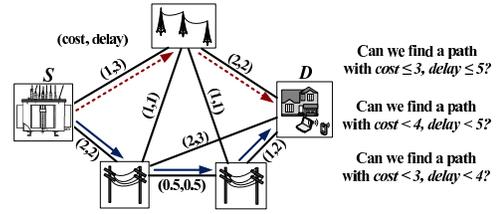}
\caption{\footnotesize Illustration of the MCR problem}
\vspace{-0.1in}
\label{fig:MCR}
\end{figure}



To see the incidences of this problem in smart grid, as shown in Fig. \ref{fig:MCR}, one can consider that on each transmission line, there are different \emph{weights} associated with it, representing {\em the energy consumption for the transmission, edge delay, edge reliability}, etc.
In smart grid, a transmission is required to satisfy several constraints, such as delay, energy consumption, and transmission reliability.
%
%
Assume that Electric generator ($S$) needs to provide QoS transmissions to the user ($D$).
On each link, two different QoS metrics: {\em cost} and {\em delay}, are considered.
If the constraint vector $W$ is (3, 5), in other words, if users aim to find a path such that {\em cost} $\leq 3$, {\em delay} $\leq 5$, path (1, 2, 5), marked by dotted red links, is a feasible path.
For the constraint vector (4, 4.5), path (1, 3, 4, 5), marked by solid blue links, is a feasible solution.
However, there is no feasible solution in this network for constraint vector (3, 4).

The {\sf MCR} problem is known to be NP-hard \cite{Wang96}, even for the case of $K = 2$.
Although QoS routing in networks has become an active area in recent years, little work, particularly on {\em performance-guaranteed multi-constrained} QoS routing, has been done in smart grid.
Given the characteristics of smart grid, there are several unique challenges for providing multi-constrained QoS routing.
Among them, one of the biggest concerns is routing for a heterogenous system like smart grid. Various devices with different resources and capabilities, from powerful large electrical generator to resource-limited sensors, are collaborated together. Most previous performance-guarantee QoS routing schemes requires strong computing capability \cite{Guerin99}\cite{Jaffe84}\cite{Lorenz01}\cite{XueZhaTanThu08}.
However, these sophisticated schemes cannot be directly applied in smart grid due to the stringent requirements on the memory and computing capability.
Second, A smart grid is a large distributed system. Most of the time, QoS routing decision must be made locally by each device based on its local information. For example, a smart meter needs to decide whether to accept a QoS request based on its local reading and expectations. Most previous work, especially with performance guarantee, requires the globe knowledge of the network, and could not be directly applied to smart grid.
To build a scheme for diverse heterogeneous system, simple and efficient QoS routing scheme, which could be implemented by various devices, is needed.
Our goal is to find a simple and effective routing solution for smart grid.

\subsection{Effective Scheme for QoS Routing in Smart Grid}

To find simple and efficient solution that can be implement in a distributed manner, we target the problem from a different perspective.
%
Instead of studying the {\sf MCR} problem directly, let us formulate an {\em optimization} version multi-constrained QoS routing problem.

\begin{definition}[{\sf OMCR}($G, s, t, K, W$)]
\label{def:MCR}

Given an undirected network $G$=$(V,E)$, with $K$ {\em nonnegative real-valued} edge weights $\omega_k(e), 1\leq k \leq K$, associate with each edge $e \in E$; a positive vector $W = \{W_1, \ldots, W_K\}$; and a source-destination node pair $(s,t)$, {\sf MCR} seeks an $s-t$ path $p_o$ such that $\omega_k(p_o) \leq \delta_o \cdot W, 1 \leq k \leq K$, where $\delta_o$ is the smallest real number $\delta \geq 0$ such that there exists an $s-t$ path $p$ satisfying $\omega_k(p)\leq \delta \cdot W_K, 1\leq k \leq K$.
\hfill $\Box$

\end{definition}

We call $\delta_o$ the {\em optimal value} of {\sf MCR} and $p_o$ ad {\em optimal path} of {\sf MCR}.
Note that $\delta_o \leq 1$ {\bf if and only if} {\sf MCR} problem is feasible.
Since $\delta_o$ could be smaller than 1, the optimization problem {\sf OMCR} also introduces a metric to compare two feasible solutions to {\sf MCR} $-$ the one with the smaller corresponding $\delta$ value is regarded as a better solution.

A very simple $K$-approximation algorithm, named {\sf OMCR}, is presented in Algorithm \ref{alg:Greedy}.
The algorithm computes an auxiliary edge weight $\omega_A(e)$ as the maximum of all edge weights $\omega_1(e), \ldots, \omega_K(e)$ divided by $W_1, \ldots, W_K$, respectively.
It then computes a shortest path $P_A$ using this auxiliary edge weight. The path $p_M$ is guaranteed to be a $K$-approximation of {\sf OMCR}.
Note that the auxiliary edge weights can be computed locally at each node, and the shortest path can be computed using Bellman-Ford algorithm.
Therefore, our $K$-approximation algorithm can be implemented as a distributed algorithm, and can be used by existing routing protocols such as OSPF \cite{Huitema00}.

\begin{algorithm}[!ht]
\small
\caption{{\sf OMCR}$(G, s, K, \vec{W}, \vec{\omega})$}
\label{alg:Greedy}
\begin{algorithmic}[1]

\FOR {each edge $e \in E$ of $G$}
\STATE
Compute an auxiliary edge weight $\omega_A(e) = \max\limits_{1\leq k \leq K} \frac{\omega_k(e)}{W_k}$;

\ENDFOR

\STATE Compute a shortest path $P_A$ from $s$ to $t$ with the auxiliary edge weight function $\omega_A$

\end{algorithmic}
\end{algorithm}

\begin{theorem}

The path $p_A$ found by Algorithm \ref{alg:Greedy} is a $K$-approximation to {\sf OMCR}.
In other words,
\begin{center}
$\omega_k(p_A) \leq K \cdot \delta_o \cdot W_k, $~~~~$ 1\leq k \leq K$,
\end{center}
where $\delta_o$ is the optimal value of {\sf OMCR}.
\hfill $\Box$
\end{theorem}

\noindent
{\sf Proof:} Since $\delta_o$ is the optimal value of {\sf OMCR}, there exists an path $p_o$ such that $\omega_k(p_o)\leq \delta_o W_k$. This means that

\begin{equation}
\label{eqn:1}
\sum_{e\in p_o} \omega_k(e) \leq \delta_o W_k, 1\leq k \leq K
\end{equation}

(\ref{eqn:1}) can be presented as

\begin{equation}
\label{eqn:2}
\sum_{e\in p_o} \frac{\omega_k(e)}{W_k} \leq \delta_o, 1\leq k \leq K
\end{equation}

Summing (\ref{eqn:2}) over $K$ constraints, we have

\begin{equation}
\label{eqn:3}
\sum_{e\in p_o} \sum_{k=1}^K \frac{\omega_k(e)}{W_k} \leq K\cdot \delta_o
\end{equation}

Since $\omega_A(e) = \max\limits_{1\leq k \leq K} \frac{\omega_k(e)}{W_k} \leq  \sum\limits_{k=1}^K \frac{\omega_k(e)}{W_k}$, we have

\begin{equation}
\label{eqn:4}
\sum_{e\in p_o} \omega_A(e) \leq \sum_{e\in p_o} \sum_{k=1}^K \frac{\omega_k(e)}{W_k} \leq K\cdot \delta_o
\end{equation}

Since $p_A$ is the shortest path with respect to edge weight function $\omega_A$, we have $\omega_A(p_A) \leq \omega_A(p_o)$. Therefore,

\begin{equation}
\label{eqn:5}
\sum_{e\in p_A} \omega_A(e) \leq \sum_{e\in p_o} \omega_A(e) \leq K\cdot \delta_o
\end{equation}

Since $\frac{\omega(e)}{W_k} \leq \omega_A(e), 1\leq k \leq K$, we have

\begin{equation}
\label{eqn:6}
\frac{\omega_k(p_A)}{W_k} = \sum_{e\in p_A} \frac{\omega(e)}{W_k} \leq \sum_{e\in p_A} \omega_A(e) \leq K\cdot \delta_o, 1\leq k \leq K
\end{equation}

Therefore, we know that $\omega_k(p_A)\leq K\cdot \delta_o\cdot W_k (1\leq k \leq k)$, and consequently, that $p_A$ is a $K$-approximation to {\sf OMCR}.

\section{Numerical Results}\label{sec:numerical}
In this section, we use numerical simulations to demonstrate the proposed mechanism and algorithm in this paper.
\subsection{Simulation Setup}
\begin{figure}[th!]
  \centering
  \includegraphics[scale=0.5]{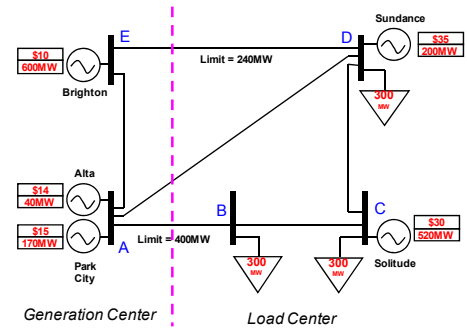}
  \caption{The base case modified from the PJM five-bus system.}\label{fig:PJM}
\end{figure}

The PJM five-bus system \cite{PJM} is used for simulations, as illustrated in Fig. \ref{fig:PJM}. The mapping between LMP and load (one curve for each bus) is given in Table \ref{tab:curve} (the first column shows the lower boundary of the corresponding load interval $\{q_i\}_{q=1,...,8}$) \cite{Li2007}.

\begin{table}[ht]
\caption{LMP (\$/MWh) versus load (MW)}
\label{tab:curve}\centering
\begin{tabular}{c c c c c c}
  \hline
  Load (MW) & LMP(A) & LMP(B) & LMP(C) & LMP(D) & LMP(E)  \\
  \hline
  0.00   & 10.00 & 10.00 & 10.00 & 10.00 & 10.00 \\
  \hline
  600.00 & 14.00 & 14.00 & 14.00 & 14.00 & 14.00 \\
  \hline
  640.00 & 15.00 & 15.00 & 15.00 & 15.00 & 15.00 \\
  \hline
  711.81 & 15.00 & 21.74 & 24.33 & 31.46 & 10.00 \\
  \hline
  742.80 & 15.83 & 23.68 & 26.70 & 35.00 & 10.00 \\
  \hline
  963.94 & 15.24 & 28.18 & 30.00 & 35.00 & 10.00 \\
  \hline
 1137.02 & 16.98 & 26.38 & 30.00 & 39.94 & 10.00 \\
  \hline
 1484.06 & 16.98 & 26.38 & 30.00 & 39.94 & 10.00 \\
  \hline
\end{tabular}
\end{table}

We assume that utility function is $U(x)=1000\log x$ and the price for communication delay is $P(d)=e^{4/d}$. Note that these functions are chosen arbitrarily for illustrative purpose. For practical systems, they can be estimated from historical data.

\subsection{QoS Requirement}
\begin{figure}[th!]
  \centering
  \includegraphics[scale=0.35]{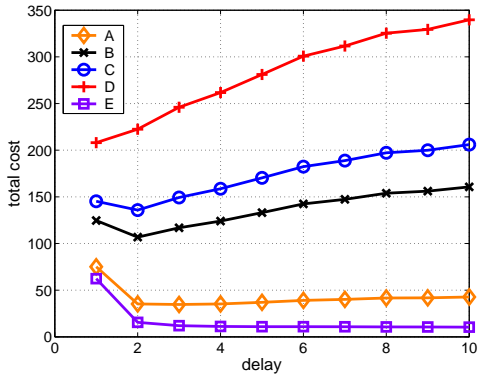}
  \caption{The curves of cost versus delay.}\label{fig:delay}
\end{figure}

\begin{figure}[th!]
  \centering
  \includegraphics[scale=0.35]{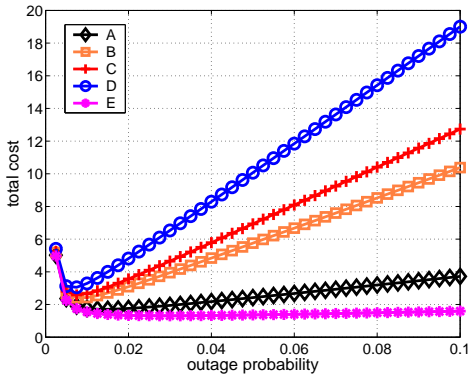}
  \caption{The curves of cost versus outage probability.}\label{fig:out}
\end{figure}

Fig. \ref{fig:delay} shows the curves of cost versus different delays (measured in time slots) for homes served by the five buses, respectively. We observe that, for some buses, the cost increases monotonically with delay while the minimal cost is not achieved by the minimal delay for other cases. Comparing the results with Table \ref{tab:curve}, we observe that, the higher the LMP is, the more sensitive the cost is to the delay. The curves of cost versus different outage probabilities are shown in Fig. \ref{fig:out}. Again, we observe the non-monotonicity of the cost, which demonstrates the existence of the optimal requirement of outage probability.

\begin{figure}[th!]
  \centering
  \includegraphics[scale=0.35]{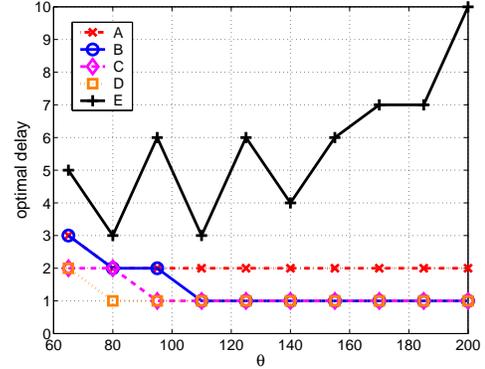}
  \caption{The optimal delay requirement when the delay and outage probability are jointed optimized.}\label{fig:opt_delay}
\end{figure}

\begin{figure}
  \centering
  \includegraphics[scale=0.35]{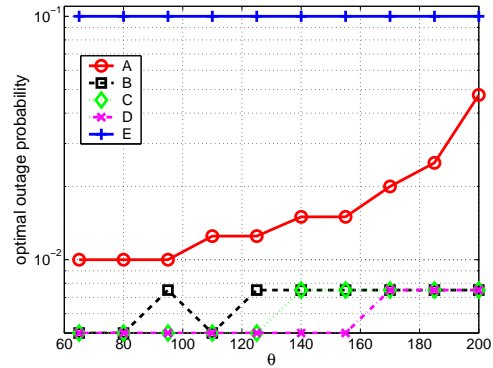}
  \caption{The optimal requirement of outage probability when the delay and outage probability are jointed optimized.}\label{fig:opt_out}
\end{figure}

The optimal requirements of delay and outage probabilities using the joint optimization in (\ref{eq:joint}) are shown in Figures \ref{fig:opt_delay} and \ref{fig:opt_out}, respectively, for various values of $\theta$. Note that the range of the outage probability is confined between 0 and 0.1. We observe that there exists some fluctuation in the optimal values. Particularly, the optimal QoS requirements of bus E are quite loose. An explanation is that the power price changes marginally for bus E. Therefore, home appliances served by bus E can degrade their QoS requirements to avoid the cost for communication.

\subsection{QoS Routing}

\begin{figure}[!th]
\centering
      \includegraphics[width=0.33\textwidth,height=0.24\textwidth]{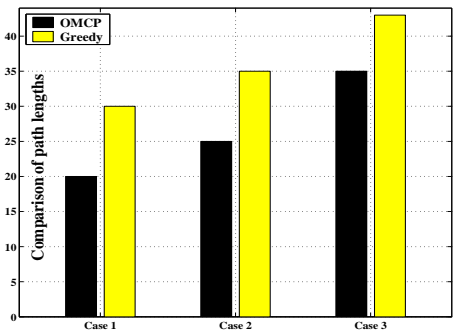}
   \caption{\footnotesize Comparison of number of better paths}
    \label{FIG03a}
\end{figure}

In this section, we present some numerical results to show the performance
of our simple greedy algorithm.
We implemented our greedy algorithm of this paper (denoted by {\sf OMCR} in
the figures), and compared it with previous sophisticated approximation
algorithm {\kFPTAS} of~\cite{XueSenZha07} (denoted by {\sf FPTAS} in the figures), which is the best approximation solution to the {\sf OMCR} problem.
Our numerical results are presented in Figs.~\ref{FIG03a}
and~\ref{FIG03c}, where each figure shows the average of $100$ runs.

First, to compare the routing performance, we define the {\em length} of a found path $p$
is $\displaystyle l(p) = \max_{1 \le k \le K}\frac{\omega_k(p)}{W_k}$.
We say path $p_1$ is better than path $p_2$ is $l(p_1) < l(p_2)$.

In Fig.~\ref{FIG03a}, we show the {\em qualitative} comparison of the
performances of {\sf OMCR} and {\sf FPTAS} using the metric of {\em path length}.
%
%
We set $\epsilon = 0.5$ for {\sf FPTAS}, which means that {\sf FPTAS} returns a 1.5-approximation to the {\sf OMCR} problem.
We observed that for all test cases, {\sf FPTAS} generally provides better results.
In among the 100 connections, in 30\% to 43\% of the test cases (30\% for the tight scenario, 35\% for the medium scenario, and 43\% for the loose scenario), the
path computed by {\sf FPTAS} is better than the path computed by {\sf OMCR}.
Meanwhile, in 20\% to 35\% of the test cases(20\% for the tight scenario, 25\% for the tight scenario, and 35\% for the loose scenario), the path computed by {\sf OMCR} is
better than the path computed by {\sf FPTAS}.
We can conclude that {\sf OMCR} has similar performance as {\sf FPTAS}.

\begin{figure}[th!]
\centering
      \includegraphics[width=0.4\textwidth,height=0.28\textwidth]{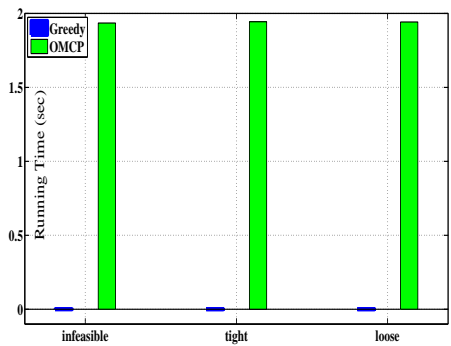}
    \vspace{-0.10in}
\caption{\small Comparison of running times}
 \label{fig:runningtime}
\end{figure}

Next, we compare the running times between the {\sf OMCR} and {\sf FPTAS} in
Fig.~\ref{fig:runningtime}.
As we expected, the running time of {\sf OMCR} is much shorter than the running
time of {\sf FPTAS}, while the two algorithms computed paths with comparable
lengths.

\begin{figure}[th!]
\centering
      \includegraphics[width=0.36\textwidth,height=0.26\textwidth]{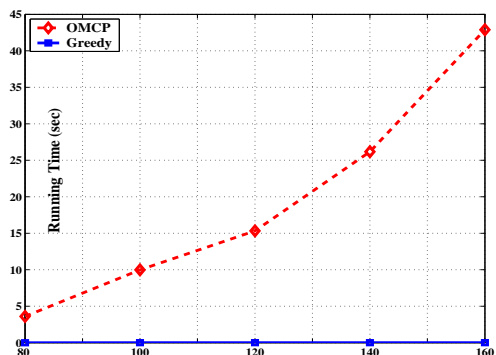}
\vspace{-0.10in}
\caption{\small Scalability of the schemes}
 \label{FIG03c}
\end{figure}

%

To study the scalability of {\sf FPTAS} and {\sf OMCR} with the network size, we used four more random network topologies with the following sizes: 80 nodes with 314 edges, 120 nodes with 474 edges, 140 nodes with 560 edges, 160 nodes with 634 edges, to test the computational scalability of the algorithms.
Here we have used $\epsilon=0.5$ and medium scenario for these test cases.
The running times of these two algorithms are shown in Fig. \ref{FIG03c}.
We can see that the running time of {\sf FPTAS} increased dramatically with the
increased network size.
Meanwhile, {\sf OMCR} requested much less amount of time and is not affect much by the size of the networks. This proves that our solution will be more adaptable in fast-developing smart grid enviroment.

\section{Conclusions}\label{sec:conclusion}
We have addressed the QoS routing in smart grid. To derive the QoS requirement, we have analyzed the dynamics of power market and the impact of communication metrics like delay and outage on the revenue of home appliances. Then, we model the QoS derivation as an optimization problem that maximizes the total reward. Based on the derived QoS requirement, a simple greedy routing algorithm has been applied to secure the QoS and address the strict realtime requirement. We have shown that the proposed algorithm is a $K$-approximation. We have run numerical simulations which demonstrated the effectiveness of the proposed mechanism and algorithm.

\section*{Acknowledgement}
The authors would like to thank Prof. Fangxing Li in the Dept. of EECS in the University of Tennessee, Knoxville, for the discussion on the load-price mapping in power market.

\bibliographystyle{IEEE}

\end{document}